\documentclass[aps,preprint,showpacs, showkeys]{revtex4}%
\usepackage{amsfonts}
\usepackage{amsmath}
\usepackage{amssymb}
\usepackage{graphicx}
\usepackage{epsfig}
\usepackage{exscale}
\usepackage{float}
\usepackage{bm}
\usepackage{suffix}
\usepackage{mathtools}
\usepackage{newunicodechar}
\usepackage{longtable}
\usepackage{lipsum}%
\usepackage{enumitem}
\usepackage{bbding}
\usepackage{tikz}
\usepackage{multirow}

\providecommand{\U}[1]{\protect\rule{.1in}{.1in}}
\DeclarePairedDelimiterX\MeijerM[3]{\lparen}{\rparen}
{\begin{smallmatrix}#1 \\ #2\end{smallmatrix}\delimsize\vert\,#3}
\newcommand\MeijerG[8][]{  G^{\,#2,#3}_{#4,#5}\MeijerM[#1]{#6}{#7}{#8}}
\WithSuffix
\newcommand\MeijerG*
[7]{
G^{\,#1,#2}_{#3,#4}\MeijerM*{#5}{#6}{#7}}

\begin{document}
\title[ ]{Weak-Coupling, Strong-Coupling and Large-Order Parametrization of the Hypergeometric-Meijer Approximants}
\author{Abouzeid M. Shalaby}
\email{amshalab@qu.edu.qa}
\affiliation{Department of Mathematics, Statistics, and Physics, Qatar University, Al
Tarfa, Doha 2713, Qatar}

\begin{abstract}
Without  Borel or Pad$\acute{e}$ techniques, we show that for a divergent series with $n!$ large-order growth factor, 
the set of Hypergeometric series $_{k+1}F_{k-1}$ represents suitable approximants for which there exist  no free parameters. The divergent $_{k+1}F_{k-1}$ series are then resummed via their representation  in terms of the Meijer-G function. The choice of $_{k+1}F_{k-1}$ accelerates the convergence even with only weak-coupling information as input. For more acceleration of the convergence, we employ the strong-coupling and large-order information. We obtained a new constraint that relates the difference of numerator and denominator parameters in the Hypergeometric approximant to one of the large-order parameters.  To test the validity of that constraint, we employed it to  obtain the exact partition function of the zero-dimensional $\phi^4$ scalar field theory.  The algorithm is also applied for the resummation of the
ground state energies of $\phi_{0+1}^{4}$ and $i\phi_{0+1}^{3}$ scalar field
theories. We get accurate results for the whole coupling space and the
precision is improved systematically in using higher orders. Precise results for the critical exponents of the $O(4)$-symmetric field model in three dimensions have been obtained from resummation of the recent six-loops order of the corresponding perturbation series.  The recent seven-loops order for the $\beta$-function of the $\phi^{4}_{3+1}$ field theory has been resummed which shows non-existence  of fixed points. The first resummation result of the seven-loop series representing the fractal dimension of  the two-dimensional self-avoiding polymer is presented here where we get a very accurate value of $d_f=1.3307$ compared to its exact value ($4/3\approx1.3333$).

\end{abstract}
\maketitle

\section{introduction}
In many situations in quantum field theory, perturbative calculations are   producing
divergent series with zero radius of convergence
\cite{zinjustin,zin-borel,Berzin,Kleinert-Borel,kleinert,kleinert2}. Being
divergent, one cannot rely on their predictions because we ignored terms that might
contribute more than the ones taken into account. To overcome such problems,
resummation techniques are introduced. The most famous one is Borel
\cite{Berzin,zinjustin} resummation technique and its extension
Borel-Pad$\acute{e}$ \cite{Kleinert-Borel}. Recently, a Hypergeometric
resummation technique has been introduced and applied to various examples \cite{Prl,cut,hyp2,hyp3,hyp4,hyp5,hyp6}. Although it results
in precise predictions in resumming a divergent series, the algorithm has some
limitations \cite{cut,Prd-GF}. As reported in Refs.\cite{cut,Prd-GF}, one might not be able to get aimed precision for
small coupling values because of the use of Hypergeometric function of finite
radius of convergence ($_{2}F_{1}$) to resum a divergent series with zero
radius of convergence. This issue has been solved (by the same
authors) for $_{2}F_{1}$ resummation in Ref.\cite{cut} by brute-force
disposition of the branch-cut (make it running from $0$ to $\infty$). Another
resummation algorithm (Borel-Hypergeometric) has been employed in Ref.\cite{Prl}
too and extended to Meijer-G approximant  algorithm in Ref.\cite{Prd-GF}. In fact, the algorithm in Ref.\cite{Prd-GF} is shown to have precise predictions from relatively low orders of perturbation series. In Ref.\cite{Alvarez}, a closely related algorithm has been used where the authors match the Borel-transformed series by a linear combination of asymptotic series of confluent Hypergeometric functions.
These algorithms can overcome the problem of precision at small coupling values. For instance, the series expansion of the used Meijer-G functions \cite{Prd-GF,G2,G3} has zero-radius of convergence while for the work in \cite{Alvarez} they are matching a Borel-transformed series with   confluent Hypergeometric functions which are in turn having finite  radius of convergence. 

The  Hypergeometric-Borel algorithm in Ref.{\cite{Prd-GF} used Pad$\acute{e}$ as well as Borel techniques to accomplish  final approximants in terms of the Meijer-G function. To apply Borel transformation to a divergent series, one needs to know the large-order growth factor ($n!$ for instance) of the given perturbation series. As long as the large-order behavior is indispensable for the application of Borel transformation, one might wonder if the Borel transformation is really needed   to achieve the Meijer-G function approximants. Besides, it is traditionally known that the incorporation of parameters from asymptotic behaviors  (strong-coupling and large-order)  of the perturbation series accelerates the convergence of resummation algorithms \cite{Kleinert-Borel} and one might in a need to suggest a way to incorporate them in the Meijer-G function parametrization. In this work we aim to introduce a resummation algorithm that incorporates information from asymptotic behaviors without using Borel or pad$\acute{e}$ techniques. The suggested algorithm   has the same level of simplicity as the first algorithm in Ref.\cite{Prl} (Hypergeometric resummation one) but on the other hand can give precise results for the whole coupling space. By simple we mean no usage of Borel or Pad$\acute{e}$ techniques but rather using Hypergeometric functions  that have zero-radius of convergence to approximate the given series and then resum them using a representation in terms of Mellin-Barnes integrals. The suggested algorithm will not only stress simplicity but also can guarantee faster convergence as it will be able to accommodate available information from asymptotic large-order and strong coupling data for the first time in such type of algorithms.

The key point to achieve our goal is to approximate the given divergent series by  the set of Hypergeometric
functions $_{p}F_{q}$ which have  zero-radius of convergence for $p\geq q+2$
\cite{HTF}. Note that$_{\text{ }k+1}F_{k}$ approximants with finite radius of convergence are still  suitable in resumming divergent
series of finite radius of convergence like the strong coupling expansion
series of the Yang-Lee model in Ref.\cite{zin-borel}. However, when the series
under consideration has a zero radius of convergence, it would be more
suitable to use the $_{\text{ }p}F_{q}$  series  with $p\geq q+2$ to approximate the divergent series under investigation.
For $p\geq q+2$, the series expansion of $_{\text{ }p}F_{q}$ is divergent but
it can be analytically continued via use of the Meijer-G function \cite{HTF}
where we have the representation:%
\begin{equation}
_{\text{ }p}F_{q}(a_{1},...a_{p};b_{1}....b_{q};z)=\frac{\prod_{k=1}^{q}%
\Gamma\left(  b_{k}\right)  }{\prod_{k=1}^{p}\Gamma\left(  a_{k}\right)  }%
\MeijerG*{1}{p}{p}{q+1}{1-a_{1}, \dots,1-a_{p}}{0,1-b_{1}, \dots, 1-b_{q}}{z}.
\label{hyp-G-C}%
\end{equation}
The Meijer-G function in turns has the integral form \cite{HTF}:
\begin{equation}
\MeijerG*{m}{n}{p}{q}{c_{1}, \dots,c_{p}}{d_{1}, \dots, d_{q}}{z} =\frac{1}{2\pi
i}\int_{C}\frac{\prod_{k=1}^{n}\Gamma\left(  s-c_{k}+1\right)  \prod_{k=1}%
^{m}\Gamma\left(  d_{k}-s\right)  }{\prod_{k=n+1}^{p}\Gamma\left(
-s+c_{k}\right)  \prod_{k=m+1}^{q}\Gamma\left(  s-d_{k}+1\right)  }z^{s}ds.
\label{hyp-G-C2}%
\end{equation}
A suitable choice of the contour $C$ enables one to get an analytic
continuation for $_{\text{ }p}F_{q}(a_{1},...a_{p};b_{1}....b_{q};z)$. For
instance when $C$ is taken from $-i\infty$ to $+i\infty$ \cite{HTF}, the
integral above converges for $p+q<2(m+n)$. For reasons that will be clearer
later, we are interested in the functions $_{\text{ }k+1}F_{k-1}%
(a_{1},...a_{k+1};b_{1}....b_{k-1};z)$ in our work. So in using
Eq.(\ref{hyp-G-C2}) we have $m=1,n=k+1$ and thus we have $p+q=2k-1$ which is is
smaller than $2\left(  m+n\right)  =2k+2$. So the resummation of the series of
$_{\text{ }k+1}F_{k-1}(a_{1},...a_{p};b_{1}....b_{q};z)$ is possible. Although
here no Borel transform is used,   the Mellin-Barnes transform defining the
G-function might suffer from Stokes phenomena \cite{Stokes} which is then
equivalent to Non-Borel summability. There exits algorithms in  literature
\cite{Stokes} to smooth them out but it is out of the scope of this work.
Instead when facing such problems, we will apply the Hypergeometric-Meijer resummation algorithm (introduced in this work) to resum the resurgent transsries \cite{Prd-GF,Stokes,Stokes1,Stokes2,instanton} associated with that
problem. The example of the resummation of the non-Borel summable series representing the zero-dimensional partition function of the degenerate-vcua $\phi^{4}$ scalar field theory will be given. 

The structure of this paper will be as follows. In Sec.\ref{LO}, we stress the strong-coupling and the large-order asymptotic behaviors of the expansion of the Hypergeometric function $_{\text{ }k+1}F_{k-1}(a_{1},...a_{p};b_{1}....b_{q};z)$. In Sec.\ref{algorithm}, the Hypergeometric-Meijer resummation algorithm is presented.  Resummation  of the divergent series of the zero-dimensional partition function of the single-vacuum (Borel-summable) and the double vacua (non-Borel summable) $\phi^4$ theory is presented in Sec.\ref{part}. In Sec.\ref{phi4} and Sec.\ref{phi3}, we apply  the resummation algorithm to the series of vacuum energies of the $\phi^{4}_{0+1}$ and the  $\mathcal{PT}$-symmetric $i\phi^{3}_{0+1}$ field theories.  The resummation results for the recent six-loops order of the renormalization group functions of the the  $O(4)$-symmetric quantum field model in three dimensions is introduced in Sec.\ref{exponent} while the application of the algorithm to resum the recent seven-loops order of the $\beta$-function of the $\phi^{4}_{3+1}$ theory is included in Sec. \ref{beta}. In Sec.\ref{fractal} , we present the first resummation result of the seven-loop ($\varepsilon$-expansion) for the fractal dimension  of the self-avoiding polymer.  Summary and conclusions will follow in Sec. \ref{conc}. 

\section{large-order and strong-coupling asymptotic behaviors of the Hypergeometric $_{k+1}F_{k-1}$  functions}\label{LO}
 
We mentioned above that toward the resummation of a divergent series with zero
radius of convergence, the functions $_{\text{ }k+1}F_{k-1}(a_{1}%
,...a_{p};b_{1}....b_{q};z)$ are suitable when the weak-coupling information
are available up to some order. It is well known that employing
strong-coupling as well as large-order data can accelerate the convergence of
a resummation technique \cite{Kleinert-Borel}. Now we need to show that the set of $_{\text{ }%
p}F_{p-2}(a_{1},...a_{p};b_{1}....b_{p-2};z)$ functions are able to
incorporate both strong-coupling as well as the large-order data of the
perturbation series to be resummed. To do that, consider the divergent series
of the expansion of a physical quantity $Q\left(  g\right)  $ such that:
\begin{equation}
Q\left(  g\right)  =\sum_{n=0}^{\infty}\beta_{n}g^{n}. \label{PS}%
\end{equation}
In fact, for divergent series of the renormalization group functions in quantum
field theory (for instance), the large-order asymptotic behavior of the perturbation series takes the
form: \cite{Kleinert-Borel}
\begin{equation}
\beta_{n}\sim\gamma n!(-\sigma)^{n}n^{b}\left(  1+O\left(  \frac{1}{n}\right)
\right)  ,\text{ \ \ }n\rightarrow\infty. \label{Large-Order}%
\end{equation}
For the resummation of a divergent series that has such kind of large order
behavior, we suggest the use of a Hypergeometric function $_{\text{ }p}F_{q}$
with a constraint on the relation between the number $p$ of numerator
parameters and the number $q$ of denominator parameters such that it can
reproduce the above large order behavior. To elucidate that point, consider the   the series expansion of the Hypergeometric function
$_{\text{ }p}F_{q}$ of the form:
\begin{align}
_{\text{ }p}F_{q}\left(  {a_{1},......a_{p};b_{1},........b_{q};-\sigma
g}\right)    & =\sum_{n=0}^{\infty}\frac{\frac{\Gamma\left(  a_{1}+n\right)
}{\Gamma\left(  a_{1}\right)  }....\frac{\Gamma\left(  a_{p}+n\right)
}{\Gamma\left(  a_{p}\right)  }}{n!\frac{\Gamma\left(  b_{1}+n\right)
}{\Gamma\left(  b_{1}\right)  }....\frac{\Gamma\left(  b_{q}+n\right)
}{\Gamma\left(  b_{q}\right)  }}\left(  -\sigma g\right)  ^{n},\nonumber\\
& =\alpha\sum_{n=0}^{\infty}\frac{\Gamma\left(  a_{1}+n\right)  ....\Gamma
\left(  a_{p}+n\right)  }{\Gamma\left(  n+1)\right)  \Gamma\left(
b_{1}+n\right)  ....\Gamma\left(  b_{q}+n\right)  }\left(  -\sigma g\right)
^{n},\nonumber\\
& =\alpha\sum_{n=0}^{\infty}c_{n}g^{n}%
\end{align}
where
\[
\alpha=\frac{%
%TCIMACRO{\dprod \limits_{i=1}^{p-2}}%
%BeginExpansion
{\displaystyle\prod\limits_{i=1}^{p}}
%EndExpansion
\Gamma\left(  b_{i}\right)  }{%
%TCIMACRO{\dprod \limits_{i=1}^{p}}%
%BeginExpansion
{\displaystyle\prod\limits_{i=1}^{p}}
%EndExpansion
\Gamma\left(  a_{i}\right)  }.
\]
 For large $n,$ the asymptotic form of a ratio of $\Gamma$ functions
is given by \cite{Gamma}:%

\begin{equation}
\frac{\Gamma\left(  n+\alpha\right)  }{\Gamma\left(  n+\beta\right)
}=n^{\alpha-\beta}\left(  1+\frac{\left(  \alpha-\beta\right)  \left(
-1+\alpha+\beta\right)  }{n}+O\left(  \frac{1}{n^{2}}\right)  \right)  .
\end{equation}
Thus for $n\rightarrow\infty$ we have:
\begin{align*}
\frac{\Gamma\left(  a_{1}+n\right)  \Gamma\left(  a_{2}+n\right)  \ }%
{\Gamma\left(  n+1\right)  \ \ }  & =\frac{\Gamma\left(  a_{1}+n\right)
\Gamma\left(  a_{2}+n\right)  \ }{\Gamma\left(  n+1\ \right)  \Gamma\left(
n+1\ \right)  }\Gamma\left(  n+1\ \right) \\& \sim n!n^{a_{1}+a_{2}-2}\left(
1+O\left(  \frac{1}{n}\right)  \right).
\end{align*}
This relation can be generalized to any number $p$ of numerator parameters and any number $q$ fo denominator parameters ($p=q+2$) as 
\begin{align*}
\frac{\Gamma\left(  a_{3}+n\right)  \Gamma\left(  a_{4}+n\right)
......\Gamma\left(  a_{p}+n\right)  \ }{\ \ \Gamma\left(  b_{1}+n\right)
.\Gamma\left(  b_{2}+n\right)  ...\Gamma\left(  b_{p-2}+n\right)  }&   \sim
n^{a_{3}+a_{4}+.......a_{p}-b_{1}-b_{2}-......b_{p-2}}\left(  1+O\left(
\frac{1}{n}\right)  \right)  .
\end{align*}
Accordingly, for the function%
\[
_{\text{ }p}F_{p-2}\left(  {a_{1},......a_{p};b_{1},........b_{p-2};-\sigma
g}\right)  =\alpha\sum_{n=0}^{\infty}c_{n}g^{n},
\]
the expansion coefficient $c_{n}$ takes the asymptotic form :%
\[
c_{n}\sim n!n^{\left(  \sum_{i=1}^{p}a_{i}-\sum_{i=1}^{p-2}b_{i}-2\right)
}\left(  1+O\left(  \frac{1}{n}\right)  \right)  ,\text{ \ \ }n\rightarrow
\infty.
\]
Thus the large-order behavior in Eq.(\ref{Large-Order}) can be reproduced from
this expansion only when $p=q+2$. Any other relation between $p$ and $q$ can not account for the $n!$ growth factor in the large-order behavior of the given divergent series. Knowing this, the large-order information  in Eq.(\ref{Large-Order}) thus sets the constraint%
\begin{equation}\label{constrain}
\sum_{i=1}^{p}a_{i}-\sum_{i=1}^{p-2}b_{i}-2=b,
\end{equation}
on the numerator parameters $a_{i}$ and the denominator parameters $b_{i}$ of
the Hypergeometric approximant $_{\text{ }p}F_{p-2}\left(  {a_{1}%
,......a_{p};b_{1},........b_{p-2};-\sigma g}\right)  $.    Accordingly, the
suitable candidate to represent the perturbation series in Eq.(\ref{PS}) with
the large order behavior in Eq.(\ref{Large-Order}) is the function $_{\text{
}p}F_{p-2}\left(  {a_{1},......a_{p};b_{1},........b_{p-2};-\sigma g}\right)
$.

The strong coupling expansion of a physical quantity can also be obtained (for  quantum field theory, it can only be obtained for
some cases) using methods in Refs.\cite{bend-strong, keinert-strong}. The
$a_{i}$ parameters in the function  $_{\text{ }p}F_{p-2}\left(  {a_{1},......a_{p}%
;b_{1},........b_{p-2};-\sigma g}\right)  $ are totally determined from powers
in the strong coupling expansion. For  non-integer $a_i-a_j$, the Hypergeometric function has the strong coupling expansion in the form \cite{mathmatica}:
	\[\, _pF_{p-2}\left(a_1,a_p;b_1,b_{p-2};z\right)\propto \sum _{k=1}^p d_k  (-z)^{-a_k}\left(1+O\left(\left(\frac{1}{z}\right)\right)\right) 
\]
From this expansion one concludes that the numerator parameters $a_i$ can be obtained from the strong coupling asymptotic behavior of the perturbation series.  So one can get the whole set of parameters in $_{\text{ }p}F_{p-2}\left(  {a_{1},......a_{p};b_{1},........b_{p-2}%
;-\sigma g}\right)  $ from the available orders of the perturbation series (weak-coupling), large-order and strong-coupling information. This means that the algorithm here does not include any free parameters like the Borel resummation with conformal mapping algorithm  in Ref.\cite{ON17} (for instance).

Sometimes one can find cases for which the differences between numerator parameters  ($a_i-a_j$) are integers and thus one might conclude that the strong-coupling asymptotic behavior of the given series can not be reproduced by any parametrization of the Hypergeometric approximants. However,  if the strong-coupling expansion of the approximant is alternating in sign, one can still extract the values of the parameters $a_{1},a_{2}......$ and $a_{p}$ from the powers in the strong coupling expansion of the given series. We shall stress this point in Sec.\ref{phi4} when studying the resummation of the ground state energy of the $\phi^4$ theory in $0+1$ dimensions.

\section{The Hypergeometric-Meijer resummation algorithm}\label{algorithm}
For a divergent series that has a large-order  $n!$ growth factor, the  Hypergeometric-Meijer resummation algorithm follows the following steps:
\begin{enumerate}
\item \textbf{Matching the given perturbation series with the series expansion of the Hypergeometric approximant \boldmath$_{\text{ }p}%
F_{p-2}(a_{1},......a_{p};b,........b_{p-2};\sigma g)$}:\\ 
 In case we have only week coupling information, all the parameters in $_{\text{ }p}%
F_{p-2}(a_{1},......a_{p};b,........b_{p-2};-\sigma g)$ are obtained by
matching the expansion of $_{\text{ }p}F_{p-2}$ with the available perturbative terms in the perturbation series  in Eq.(\ref{PS}).
For example, the third order parametrization will lead to the $_{2}F_{0}%
(a_{1},a_{2};\ ;-\sigma g)$ where the matching will lead to the result:
\begin{align}
a_{1}a_{2}\sigma &  =\beta_{1}\nonumber\\
\frac{1}{2}a_{1}\left(  1+a_{1}\right)  a_{2}\left(  1+a_{2}\right)
\sigma^{2} &  =\beta_{2}\label{w-algo}\\
\frac{1}{6}a_{1}\left(  1+a_{1}\right)  \left(  2+a_{1}\right)  a_{2}\left(
1+a_{2}\right)  \left(  2+a_{2}\right)  \sigma^{3} &  =\beta_{3}\nonumber
\end{align}
Solving these equations, the three parameters $a_{1}, a_{2}$ and $\sigma$ are
fully determined. The fourth order Hypergeometric approximant is $_{3}%
F_{1}(a_{1},a_{2},a_{3};\ b_{1};-\sigma g)$ and so on.

\item \textbf{Hypergeometric to Meijer-G approximants}: We use the representation of the  Hypergeometric function $_{\text{ }p}F_{p-2}(a_{1}%
,......a_{p};b,........b_{p-2};-\sigma g)$ in terms of the Meijer-G function
in Eq.(\ref{hyp-G-C}) to get a convergent result out of the divergent series for
\ $_{\text{ }p}F_{p-2}.$ For instance, the fourth order Hypergeometric approximation is represented as:

\begin{equation}
_{\text{ }3}F_{1}(a_{1},a_{2},a_{3};b_1 ;\sigma z)=\frac{
\Gamma\left(  b_{1}\right)  }{\prod_{k=1}^{3}\Gamma\left(  a_{k}\right)  }%
\MeijerG*{1}{3}{3}{2}{1-a_{1}, \dots,1-a_{3}}{0,1-b_{1}}{ \sigma z}.
\end{equation} 
\end{enumerate}

One can accelerate the convergence of the algorithm by using the large-order
information. To illustrate this, consider for  simplicity the $_{2}%
F_{0}(a_{1},a_{2};\ ;-\sigma g)$ which needs three orders from the perturbation
series represented by the coefficients $\beta_{1}$, $\beta_{2\text{ }}$ and $\beta_{3}$
above. In case we know the large-order information, one can match the parameters known from the large-order behavior of the given perturbation series  with the large order form of  $_{2}F_{0}(a_{1},a_{2};\ ;-\sigma g)$. Thus the parameter $\sigma$ is known from that large-order behavior and also     the other parameters are constrained as:
\begin{equation} 
\sum_{i=1}^{p}a_{i}-\sum_{i=1}^{p-2}b_{i}-2=b. \label{LOB}%
\end{equation}
Accordingly, the large-order information lowers the third order parametrization of $_{2}F_{0}(a_{1},a_{2};\ ;-\sigma g)$  to just first-order. This means that we need only the equation $a_{1}a_{2}\sigma=\beta_{1}$ from weak-coupling data and the constraint in Eq.(\ref{LOB}) to solve for $a_1$ and $a_2$. A note to be mentioned on using low-order approximants is that, except for rare cases, one usually can not extract good approximations from just first order of perturbation series  as input and of course good approximations are always expected for second , third and higher orders.

 In case we  know the strong-coupling information, then the parameters  $a_{1}$,
$a_{2}$ are known \cite{Abo-hyper} while $\sigma$  is already known form large-order data. In other words,   all the parameters in $_{2}F_{0}(a_{1},a_{2};\ ;-\sigma g)$   have been determined completely without the need of the week
coupling information. In other words, employing the strong coupling information beside the weak coupling can accelerate the convergence to the extent that we are not in a need to weak-coupling data for the lowest order approximant:
\[
_{\text{ }2}F_{0}\left(  {a_{1},a_{2};\ ;\sigma g}\right)  =\frac{1}%
{\Gamma\left(  a_{1}\right)  \Gamma\left(  a_{2}\right)  }\MeijerG*{1}{2}{2}{1}{1-a_{1},1-a_{2}}{0}{ \sigma z}. 
\]
Again we need to assert that in using weak-coupling, strong-coupling and large-order parametrization, the lowest order approximant  $_{2}F_{0}(a_{1},a_{2};\ ;-\sigma g)$   is not always expected to produce accurate results and to get good approximations, one resorts to higher order approximants.

 For  a  higher orders Hypergeometric function $_{\text{ }p}F_{p-2}(a_{1},......a_{p}%
;b,........b_{p-2};-\sigma g)$ with their equivalent Meijer-G approximants, one needs $2p-1$ terms from the
week-coupling information to solve for all unknown parameters $a_i$,$b_i$ and $\sigma$. If the strong coupling information are known, we
need only $p-1$ terms from week-coupling information. Knowing  
 weak-coupling, strong-coupling and large-order information, then one needs $p-3$ orders of the
perturbation series to determine all the parameters in the $_{\text{ }%
p}F_{p-2}(a_{1},......a_{p};b,........b_{p-2};-\sigma g)$ series.

\section{Hypergeometric-Meijer resummation of   zero-dimensional partition function of the $\phi^4$ scalar field theory}\label{part}
In this section we give two examples for resummation of the partition function of $\phi^4$ theory in zero dimension, where it has a divergent series expansion. The first case is  the single vacuum  theory where the series is Borel summable and no complex ambiguity exists. The second example is the partition function of a double-vacua $\phi^4$ theory where the series is non-Borel summable and thus one resorts to the resummation of Resurgent Transseries.
\subsection{ Single-vacuum $\phi^4$  Scalar Field Theory}

An example of a divergent series with zero radius of convergence that is always used to test the success of a resummation algorithm is the partition function of zero-dimensional $\phi^4$ theory. Let us apply the algorithm here to resum the associated divergent perturbation series. We shall apply the algorithm three times for the same problem, one using weak-coupling information only, another by  adding the large-order information and finally by adding strong-coupling information. The reason behind using  that recipe is to test the validity of the new constraint set on the parameters in Eq.(\ref{constrain}) using    an exact  resummation result. To do that, consider  the partition function of that model given by:%
\begin{equation}
Z=\frac{1}{\sqrt{2\pi}}\int_{-\infty}^{\infty}d\phi\exp\left(-\frac{\phi^{2}%
}{2}-\frac{g}{4!}\phi^{4}\right)  \label{part-zd0},
\end{equation}
whith the associated weak-coupling perturbation series is of the form:
\begin{equation}
Z\left(  g\right)  =1-\frac{g}{8}+\frac{35}{384}g^{2}-\frac{385}{3072}%
g^{3}+O\left(  g^{4}\right)  \label{par-d0}.
\end{equation}
In fact, the lowest order  $_{\text{ }p}F_{p-2}\left(  a_{1},......a_{p}%
;b,........b_{p-t};\sigma g\right)  $ Hypergeometric approximant is $_{\text{ }%
2}F_{0}\left(  a_{1},a_{2};\text{ \ };\sigma g\right)  $ with only three
unknown parameters. To determine the parameters $a_{1}$, $a_{2}$ and $\sigma$
we use Eq.(\ref{w-algo}) with the corresponding $\beta_i$ coefficients:
\begin{align*}
a_{1}a_{2}\sigma &  =-\frac{1}{8}\\
\frac{1}{2}a_{1}\left(  1+a_{1}\right)  a_{2}\left(  1+a_{2}\right)
\sigma^{2}  &  =\frac{35}{384}\\
\frac{1}{6}a_{1}\left(  1+a_{1}\right)  \left(  2+a_{1}\right)  a_{2}\left(
1+a_{2}\right)  \left(  2+a_{2}\right)  \sigma^{3}  &  =-\frac{385}{3072}.
\end{align*}
The solution of these equations are given by: $a_{1}=\frac{1}{4},a_{2}%
=\frac{3}{4}$ and $\sigma=-\frac{2}{3}$. Accordingly, the Hypergeometric-Meijer
approximant of $Z\left(  g\right)  $ is
\begin{align*}
Z\left(  g\right)    & =_{\text{ }2}F_{0}\left(  \frac{1}{4},\frac{3}%
{4};\text{ \ };-\frac{2}{3}g\right)  \\
& =\frac{
1 }{\Gamma\left(\frac{1}{4}\right) \Gamma\left( \frac{3}{4}\right) }%
\MeijerG*{1}{2}{2}{1}{\frac{3}{4}, \frac{1}{4}}{0}{ \frac{2}{3} g}
\end{align*}
In using the identity (see Eq.(9) in sec.5.3.1 and Eq.(7) in sec.5.6 of Ref.\cite{HTF})
\[
\,_{2}F_{0}(-n,n+1,x)=\frac{1}{\sqrt{\pi}}\sqrt{\frac{-1}{x}}\exp\left(
-\frac{1}{2x}\right)  K_{n+\frac{1}{2}}\left(  -\frac{1}{2x}\right)  ,
\]
we get the exact result reported in Ref.\cite{Prd-GF} but it has been obtained
there at the fifth order while we obtained it from knowing only the first three
terms of the week coupling expansion. 

One can even accelerate  the convergence  to the exact result by using the large-order information. The large-order behavior for
the series $Z\left(  g\right)  $ can also be obtained as $n!n^{-1}\left(
\frac{-2}{3}g\right )  ^{n}$ $\left(  1+O\left(  \frac{1}{n}\right)  \right)  $
for $n\rightarrow\infty$. Accordingly, we have  $\sigma=-\frac{2}{3}.$ For the parameters
$a_{1}\ $and $a_{2}\ $, we use one equation from matching the weak-coupling
expansion with expansion of \ $_{\text{ }2}F_{0}\left(  a_{1},a_{2};\text{
\ };\sigma g\right)$ to get:

\[
\frac{-2}{3}a_{1}a_{2}=-\frac{1}{8},
\]
while the other equation from matching the large-order behavior  in
Eq.(\ref{LOB}):
\[
a_{1}+a_{2}-2=-1.
\]
Solving these equations one gets: $a_{1}=\frac{3}{4}$ and $a_{2}=\frac{1}{4}$. So in using the large-order data, the exact result has been obtained from first order in perturbation series. This result assures the validity of the new constraint   obtained in this work (Eq.(\ref{constrain})).

One can also make the convergence even faster in case we know also the strong-coupling information. The strong
coupling expansion of the integral in Eq.(\ref{part-zd0}) can be obtained as:
\begin{equation}
Z\left(  g\right)  =\frac{\sqrt[4]{\frac{24}{16}}\sqrt{\pi}}{\Gamma\left(
\frac{3}{4}\right)  }g^{-\frac{1}{4}}-\sqrt[4]{\frac{3}{2}}\frac{\Gamma\left(
\frac{3}{4}\right)  }{\sqrt{\frac{\pi}{3}}}g^{-\frac{3}{4}}\ +O\left(
g^{-\frac{5}{4}}\right)  . \label{part-strong}
\end{equation}
When $a_{1}-a_{2}$ is not an integer, the asymptotic behavior of \ $_{\text{
}2}F_{0}\left(  a_{1},a_{2};\text{ \ };\sigma g\right)  $ for large $g$ values
takes the form:%
\[
_{\text{ }2}F_{0}\left(  a_{1},a_{2};\text{ \ };\sigma g\right)  \sim
c_{1}g^{-a_{1}}+c_{2}g^{-a_{2}}.
\]
Accordingly, we get $a_{1}=\frac{3}{4}$ and $a_{2}=\frac{1}{4}$. Thus we know
$\sigma$ from large-order data and $a_{1}$and $a_{2}$ from strong-coupling
data. Thus the exact partition function has been obtained from the
knowledge of the large-order and strong-coupling information only (no
week-coupling data needed). 
\subsection{  Double-Vacua $\phi^4$ Theory}
In some cases, the perturbation series is not Borel-summable and the Borel-summation of the series leads to complex ambiguities \cite{Stokes1}. An example of such kind of perturbation series is the one associated with the integral representing the zero-dimensional partition function of the degenerate-vacua $\phi^4$ theory \cite{Prd-GF,Stokes,Stokes2}:
\begin{equation}
Z=\frac{1}{\sqrt{2\pi}}\int_{-\infty}^{\infty}d\phi\exp\left( - \frac{\phi^{2}%
}{2}\left(  1-\sqrt{g}\phi^{2}\right)  ^{2}\right), 
\end{equation}
where  it has an expansion of the from:
\[
Z\left(  g\right)  \sim1+6g+210g^{2}+13860g^{3}+O\left(  g^{4}\right)  .
\]
It is clear that this series is not Borel-summable and the Borel sum will
result in a complex ambiguity. The reason behind that is the existence of singular points on the contour used in the Borel transform which results in the existence of Stokes phenomena. A similar situation can exist for the Meijer-G resummation since the   Meijer-G function  is defined through a Mellin-Barnes integrals where Stokes phenomena can exist too  \cite{Stokes2}. In such case a resurgent transseries can be obtained that can account for non-perturbative contributions for small coupling values associated with the expansion around the non-perurbative saddle point   \cite{Stokes1,Stokes,Stokes2}.  
The transseries for the zero-dimensional partition function of the degenerate-vacua $\phi^4$ theory has been reported in  Ref.\cite{Prd-GF} as: 
\begin{align}
Z\left(  g\right)    & =\pm i\sqrt{2}\exp\left(  \frac{-1}{32g}\left(
1-6g+210g^{2}-13860g^{3}+O\left(  g^{4}\right)  \right)  \right)  \nonumber\\
& +2\left(  \left(  1+6g+210g^{2}+13860g^{3}+O\left(  g^{4}\right)  \right)
\right),
\end{align}
where the $+$ sign for $Im(g)>0$ and $-$ sign for $Im(g)<0$. This transseries has in fact incorporated the contributions from the Gaussian saddle point and the instanton saddle point \cite{instanton}. The two separate series in the transseries above can be resummed using the Hypergeometric-Meijer  Resummation followed in this work and the exact result is obtained at the third order where we have:
\begin{align*}
Z\left(  g\right)    & =\frac{
2}{\prod_{k=1}^{2}\Gamma\left(  a_{k}\right)  }%
\MeijerG*{1}{2}{2}{1}{1-a_{1}, 1-a_{2}}{0}{ -32 g}+\frac{ \pm
i \sqrt{2} e^{\left.-\frac{1}{32}\right/g}}{\prod_{k=1}^{2}\Gamma\left(  a_{k}\right)  }%
\MeijerG*{1}{2}{2}{1}{1-a_{1}, 1-a_{2}}{0}{ 32 g},
\end{align*}
with  $a_{1}=\frac{1}{4}$ and  $a_{1}=\frac{3}{4}$. Note that this result is real and exact. In Ref.\cite{Exact-z}, the exact result is listed as (for $Re(g)>0$):

	\[ Z(g)=\frac{e^{\left.-\frac{1}{64}\right/g} D_{-\frac{1}{2}}\left(-\frac{1}{4 \sqrt{g}}\right)}{\sqrt{2} \sqrt[4]{g}},
\]
where $D_{\nu}(z)$ is the parabolic cylinder function. The Meijer G-function approximants above gives for  $Z(2)=0.778225$ , $Z(20)=0.417229$ and $z(200)=0.230612$ which are the exact numerical values from the  parabolic cylinder function above. 

 One can reduce the order to first order only in using the large-order information and to zero order in using strong-coupling information.

\section{Resummation of the vacuum energy perturbation series of the $\phi^{4}_{0+1}$ Scalar field theory}\label{phi4}
As another testing example, we apply the algorithm to resum the ground-state energy of the anharmonic
oscillator where it is equivalent to the scalar $\phi^{4}$ theory in $0+1$
space-time dimensions. We shall resum the same series using two different parametrizations. The first parametrization is using weak-coupling, large-order and strong-coupling data. In the second parametrization, we use weak-coupling and large-order data while the strong-coupling parameters  are extracted from the approximant. Up to the best of our knowledge, a closed form  strong-coupling asymptotic behavior has not been obtained yet even for simple quantum field theories like the $\phi^4$-scalar field theory in space-time dimensions higher than $0+1$. Accordingly, the second parametrization is very important in obtaining the asymptotic strong-coupling behavior in quantum field theories where other resummation algorithms can give different results for the same problem \cite{BST1,BST2,BST3,BST4}. 
\subsection{Weak-coupling, large-order and strong-coupling  parametrization of resummation approximants for $\phi^{4}_{0+1}$  vacuum energy}
The Hamiltonian density for this example is given by:
\begin{equation}
H=\frac{\pi^{2}}{2}+\frac{m}{2}\phi^{2}+\frac{g}{4}\phi^{4}. \label{phi4H}%
\end{equation}
In $0+1$ space-time dimensions and for $m=1$, the perturbation series of the
ground state energy has the form \cite{benderx4-large}:%
\begin{equation}
E_{0}=\frac{1}{2}+\frac{3}{4}g-\frac{21}{8}g^{2}+\frac{333}{16}g^{3}%
-\frac{30885}{128}g^{4}+\frac{916731}{256}g^{5}+O(g^{6}). \label{x4-pert}%
\end{equation}
The large order behavior is given also by $-(-3)^{n}\sqrt{\frac{6}{\pi^{3}}%
}\Gamma\left(  n+\frac{1}{2}\right)  $ \cite{benderx4-large}. It is clear here that the parameter $\sigma$ is then given by
$\sigma=3$. A scaling operation can lead to the strong coupling expansion \cite{kleinert2} from
which one can extract $a_{i}$ as :

\[
a_{1}=-\frac{1}{3},a_{2}=\frac{1}{3},a_{3}=1,a_{4}=\frac{5}{3},......
\]
For  the approximants $_{\text{ }2}F_{0}$ and $_{\text{ }3}F_{1}$, the difference between any two numerator parameters   ($a_i-a_j$) can't be integer and thus the numerator parameters lead  to the well-known strong coupling asymptotic behavior \cite{mathmatica}.  For all higher orders approximants ($_{\text{ }4}F_{2}$, $_{\text{ }5}F_{3}$ .......),  however, $a_i-a_j$ has the possibility to take integer values and thus lead to logarithmic factors in the strong-coupling asymptotic behavior\cite{mathmatica} which does not mach with the known strong-coupling expansion of the given series. In fact, the logarithmic factors in the strong coupling asymptotic behavior are multiplied by $g^{-a_4}$ and $g^{-a_5}$ for  $_{\text{ }5}F_{3}$ (for instance) which means that such terms will be led by the power behaviors $g^{-a_4}$ and $g^{-a_5}$ while the logarithmic factors have minor effect at large $g$. This means that although  for some approximants, $a_i-a_j$ have the possibility to be integers, one can still consider the numerator parameters matching the exact ones known from strong-coupling expansion of the given series. To test   these expectations, we    parametrized   the approximant $_{\text{ }5}F_{3}$ in two ways, one by setting  $a_1,a_2,a_3,a_4$ and $a_5$ to match with known exact ones  while for the other parametrization we take $a_1,a_2,a_3$ from known exact ones while predicting $a_4$ and $a_5$ by considering more terms from the weak-coupling data. We found only marginal differences between the predictions of the two parametrization (Fig.\ref{tpara}).
\begin{figure}[t]
\begin{center}
\epsfig{file=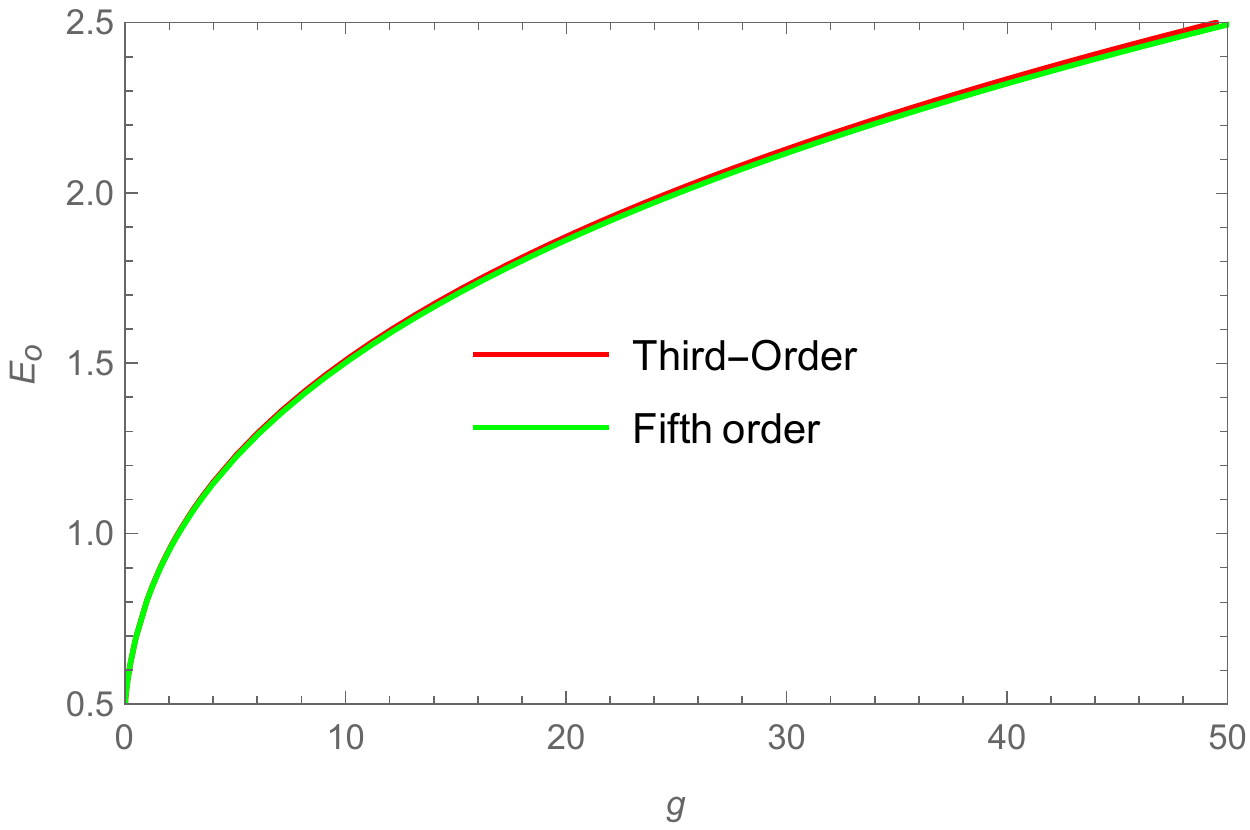,width=0.65\textwidth}
\end{center}
\caption{\textit{Comparison between resummation of the ground state energy for the $\phi^4_{0+1}$ theory using two different parametrization of the $_{\text{ }5}F_{3}$ approximant. In the third order parametrization (red color) we set all values of numerator parameters to be $-\frac{1}{3},\frac{1}{3},1,\frac{5}{3} $ and $\frac{7}{3}$ from the known strong coupling expansion of the theory. For the other parametrization (fifth order, green color), we take  $a_1=-\frac{1}{3},a_2=\frac{1}{3},a_3=1 $ while $a_4$ and $a_5$ are obtained by considering two more orders from perturbation series. }}
\label{tpara}%
\end{figure}

 A concrete advocate of the irrelevance of existing singular coefficients in the strong coupling expansion of the Hypergeometric approximants  can be introduced by more deep analysis of its properties. For the given series,   the strong coupling behavior of the approximant  $\frac{_{5}F_{3}}{2}$ (for instance) is given by: 
 
	 \[\frac{_{5}F_{3}}{2}%
\propto g^{-a_{1}}\left(
c_{1}+c_{2}g^{a_{1}-a_{2}}+c_{3}g^{a_{1}-a_{3}}+c_{4}g^{a_{1}-a_{4}}+c_{5}g^{a_{1}-a_{5}}\right) 
 \]
 Here $c_{1},c_{2}$ and $c_{3}$ are finite but$%
c_{4} $ and $c_{5}$ are singular. Let us write them explicitly:
\begin{eqnarray*}
c_{4} &=&\frac{\left( 3\right) ^{-a_{4}}\Gamma (b_{3})\Gamma (b_{2})\Gamma
(b_{1})\Gamma (a_{2}-a_{4})\Gamma (a_{3}-a_{4})\Gamma (a_{5}-a_{4})\Gamma
(a_{1}-a_{4})}{2\Gamma (a_{1})\Gamma (a_{2})\Gamma (a_{3})\Gamma
(a_{5})\Gamma (b_{1}-a_{4})\Gamma (b_{2}-a_{4})\Gamma (b_{3}-a_{4})} \\
c_{5} &=&\frac{\left( 3\right) ^{-a_{5}}\Gamma (b_{3})\Gamma (b_{2})\Gamma
(b_{1})\Gamma (a_{2}-a_{5})\Gamma (a_{3}-a_{5})\Gamma (a_{4}-a_{5})\Gamma
(a_{1}-a_{5})}{2\Gamma (a_{1})\Gamma (a_{2})\Gamma (a_{3})\Gamma
(a_{4})\Gamma (b_{1}-a_{5})\Gamma (b_{2}-a_{5})\Gamma (b_{3}-a_{5})}
\end{eqnarray*}
Clearly these coefficients are singular but also of opposite signs. One thus can
hope to regularize the fourth and the fifth terms in the
strong-coupling expansion. To do that let us substitute the given parameters
and add a fictitious variable $\varepsilon $ in the singular terms and then
take the limit as $\varepsilon \rightarrow 0$. In this case we have

\begin{eqnarray*}
c_{4}g^{-5/3}+c_{5}g^{-7/3} &=&\frac{(0.0210944\,\ )\Gamma (\varepsilon -2)}{%
g^{5/3}}-\frac{(0.0564589\,\ )\Gamma (\varepsilon -2)}{g^{7/3}} \\
&=&\frac{1}{\varepsilon }\left( \frac{1.\,\allowbreak 054\,7\times 10^{-2}}{%
g^{\frac{5}{3}}}-\frac{2.\,\allowbreak 822\,9\times 10^{-2}}{g^{\frac{7}{3}}}%
\right) \\
&&+\left( \frac{9.\,\allowbreak 732\,8\times 10^{-3}}{g^{\frac{5}{3}}}-\frac{%
2.\,\allowbreak 605\,0\times 10^{-2}}{g^{\frac{7}{3}}}\allowbreak \right)
+\allowbreak O\left( \varepsilon \right) \ 
\end{eqnarray*}%
Now taking the limit as $\left( g,\varepsilon \right) \rightarrow \left(
\infty ,0\right) $, we get%
\begin{equation*}
c_{4}g^{5/3}+c_{5}g^{7/3}=\left( \frac{9.\,\allowbreak 732\,8\times 10^{-3}}{%
g^{\frac{5}{3}}}-\frac{2.\,\allowbreak 605\,0\times 10^{-2}}{g^{\frac{7}{3}}}%
\allowbreak \right)
\end{equation*}
In this case the strong-coupling approximation for the ground state energy then takes the form:

\begin{equation*}
E_{0}\left( g\right) \approx 0.66649\,g^{\frac{1}{3}}+\frac{%
0.13506}{g^{1/3}}-\frac{0.072736}{g}+\frac{%
0.00973 }{g^{\frac{5}{3}}}
\end{equation*}
One can test the validity of this result by taking $g=50\allowbreak $ to get $E_{0}\left(
50\right) \approx 2.\,\allowbreak 490\,6$ while the full $_{5}F_{3}$
approximant gives $E_{0}=2.4936141$ \ compared to exact  result as $%
E_{0}=2.49971$. Note that strictly speaking $c_{4}$ is singular but since
the strong-coupling series is alternating in sign, singular
coefficients can be regularized in pairs. 

In going from low-order approximants to higher orders, the convergence is improved in a systematic way. For instance, at  $g=1$,  we have the resummation results as follows: for $_{2}F_{0}$ approximant we
have $E_{0}=0.599859$ ,   $_{3}F_{1}$ approximant gives $E_{0}=0.864509,$ 
while $_{4}F_{2} \ $gives $E_{0}=0.794639$ and for $_{5}F_{3}$, we get $\
E_{0}=0.803068$ \ compared to the exact   result as $E_{0}=0.8037706512$.
One can realize that the accuracy is improving from order to order in a
systematic way and this is not solely for $g=1$ but for the whole coupling
space. For instance at $g=50,$ $_{2}F_{0}$ gives \ $E_{0}=1.4270926$ ,   $%
_{3}F_{1}$   gives $E_{0}=3.047260,$ while $_{4}F_{2} \ $ gives  
$E_{0}=2.474683$ and for $_{5}F_{3}$, we get $\ E_{0}=2.4936141$ \ compared
to the exact   result as $E_{0}=2.49971$. The  zero
order approximant $_{2}F_{0}$ gives inaccurate result but this is
because no input information used from perturbative series. The first order $%
_{3}F_{1}$ approximant gives reasonable approximation specially if we know that it
uses only the first order from the perturbation series as input. At the second order $%
_{4}F_{2}$, the result has been greatly improved while the third order $%
_{5}F_{3}$ approximant gives better accuracy. so, the convergence improves by
going to higher orders in a systematic way and one should not rely on the jump 
from zero to first order resummation results as one can not expect convergence that fast for any resummation algorithm. 

As we mentioned above  $_{\text{ }3}F_{1}$, $_{\text{ }4}F_{2}$, $_{\text{ }5}F_{3}$ approximants give good results with convergence  improvement
from order to order. We also tested the $_{\text{ }6}F_{4}$ approximant and it shows better improvement. However, in the following  we will present  the details about   $_{\text{ }%
8}F_{6}$ approximant only (fifth order in using weak-coupling, large-order and strong coupling data). The $b_{i}-$parameters in the $_{\text{ }8}F_{6}$ function can then be obtained from matching the coefficients of the series expansion of
$_{\text{ }8}F_{6}$ term by term with the coefficients in the perturbation
series in Eq.(\ref{x4-pert}). We obtained the following values for the parameters $b_{i}:$
\begin{align*}
b_{1}  & =0.448491,\text{ }b_{2}=1.02679-2.64427i,\text{ }b_{3}=b_{2}^{\ast
},\\
b_{4}  & =0.585824-0.748355I,b_{5}=b_{4}^{\ast},b_{6}=12.6379.
\end{align*} 

Accordingly, the fifth order resummation gives;
	\begin{equation}
E_0=\frac{1}{2} \, _8F_6\left(a_{1},...a_{8};b_{1}....b_{6};\sigma g\right)=\frac{\prod_{k=1}^{6}%
\Gamma\left(  b_{k}\right)  }{2\prod_{k=1}^{8}\Gamma\left(  a_{k}\right)  }%
\MeijerG*{1}{8}{8}{7}{1-a_{1}, \dots,1-a_{8}}{0,1-b_{1}, \dots, 1-b_{6}}{\sigma g}.
\end{equation}
The predictions of this order of resummation are shown in table \ref{tab:wgresum-x4}. It is very clear that the algorithm gives accurate
results from a relatively low order of the given perturbation sires.
\begin{table}[ H]
\caption{{\protect\scriptsize {The fifth order Hypergeometric-Meijer  resummation $_{8}F_{6}$ for
the ground state Energy in Eq.(\ref{x4-pert}) compared to the exact results
from Ref.\cite{Ivanov}.}}}%
\label{tab:wgresum-x4}%
\begin{center}
\begin{tabular}
[c]{|l|l|l|}\hline
g &\ \ \  $_{8}F_{6}$  &\ \  Exact\\\hline
0.5 & 0.6961203131 & 0.6961758208\\\hline
1 & 0.8037160010 & 0.8037706512\\\hline
50 & 2.500620727 & 2.4997087726\\\hline
1000 & 6.702747381 & 6.694220850 5\\\hline
20000 & 18.16565096 & 18.137229073\\\hline
\end{tabular}
\end{center}
\end{table}
\subsection{Predicting the asymptotic strong-coupling behavior using Hypergeometric-Meijer algorithm}

The fact that our algorithm doesn't  have free parameters, makes it the most suitable algorithm in extracting the strong-coupling asymptotic behavior from weak-coupling and large order data as input. The Borel with conformal mapping algorithm in Ref.\cite{ON17} (for instance) includes three free parameters that are optimized to give the best convergence.  However, it has been shown in the literature \cite{BST1,BST2,BST3,BST4} that different optimizations can lead to different strong-coupling behaviors for the same theory. In the following we extract the asymptotic strong-coupling behavior of the ground state energy of the $\phi^4_{0+1}$ theory and compare it with known exact results. 

For the approximant  $\, _4F_2\left(a_{1},...a_{4};b_{1},b_{2};\sigma g\right)$ for instance, the above discussions telling us that the strong coupling behavior is given by  $\, _4F_2\left(a_{1},...a_{4};b_{1},b_{2};\sigma g\right) \propto g^s$ where $s=Max(-a_{1},-a_{2},-a_{3},-a_{4})$. We parametrized the approximant $\, _4F_2\left(a_{1},...a_{4};b_{1},b_{2};\sigma g\right)$ for the ground state energy using weak-coupling and large-order data and found that as $g\rightarrow \infty$ we have $E_0\propto g^{s}$ with $S=0.325731$. This is a fifth order prediction for $s$ while the exact value is $s=1/3\approx0.333333$ as shown above. Of course higher order approximants shall give better prediction for $s$. Accordingly, one can claim that our algorithm can be used to predict accurate   asymptotic strong-coupling behavior of a divergent series from the knowledge of weak-coupling and large-order data. Not only that, but it can even answer a long lasting question of why for instance the first few perturbative orders of the epsilon-expansion can give accurate critical exponents \cite{Kaku} while adding more orders will ruin the accuracy. The point is that if one of the $a_i$ parameters is a negative integer $-l$, the Hypergeometric series tends to be a truncated polynomial of order $l$. We will stop at this point as these type of discussions will appear somewhere else.

\section{Vacuum energy of the $\mathcal{PT}$-symmetric $i\phi^{3}_{0+1}$ theory}\label{phi3}
Another example for a divergent series with zero radius of
convergence is the ground state energy of the $\mathcal{PT}$-symmetric $i
\phi^{3}$ theory with Hamiltonian density operator in the form:
\begin{equation}
H=\frac{1}{2}\pi^{2}+\frac{1}{2}\left(  \nabla\phi\right)  ^{2}+\frac{1}%
{2}m^{2}\phi^{2}(x)+\frac{i\sqrt{g}}{6}\phi^{3}\left(  x\right)  .
\label{iphi3}%
\end{equation}
In $0+1$ space-time dimensions, the ground state energy of this theory has  the perturbation
series \cite{benderx3-large}
\begin{equation}
E_{0}=\frac{1}{2}+\frac{11g}{288}-\frac{930}{288^{2}}g^{2}+\frac
{158836}{288^{2}}g^{3}+\frac{38501610}{288^{4}}g^{4}+O\left(  g^{5}\right).
\label{pertub}%
\end{equation}
 Also, the strong coupling parameters are given in Ref.\cite{zin-borel} as:
\[
a_{1}=1,a_{2}=-\frac{1}{5}\ ,a_{3}=\frac{3}{5},a_{4}=\frac{7}{5},a_{5}%
=\frac{11}{5},a_{6}=\frac{15}{5}.
\]
In using these parameters and matching the expansion of $_6F_{4}$ with expansion in Eq.(\ref{pertub}), we get the numerators parameters $b_i$ as:
\begin{align*}
b_{1}  & =0.43189086698613627` - 1.2561659803549978`i,\text{ }b_{2}=b_{1}^{*},\\
 \text{ }b_{3}&=4.605221446564435 ,b_{4}   =0.3721464374405092.
\end{align*} 
Accordingly, the fourth order Hypergeometric-Meijer resummation for the vacuum energy is:

\begin{equation}
E_0=\frac{1}{2}_{\text{ }6}F_{4}(a_{1},...a_{6};b_{1}....b_{4};\sigma z)=\frac{\prod_{k=1}^{4}%
\Gamma\left(  b_{k}\right)  }{2\prod_{k=1}^{6}\Gamma\left(  a_{k}\right)  }%
\MeijerG*{1}{6}{6}{5}{1-a_{1}, \dots,1-a_{6}}{0,1-b_{1}, \dots, 1-b_{4}}{\sigma z}.
\end{equation}
The vacuum energy of the $\mathcal{PT}$-symmetric $i\phi^{3}_{0+1}$ theory has been resummed using different techniques in Ref.\cite{zin-borel}. Our calculations are shown in table-\ref{tab:wgresum} where it is  compared to the $150^{th}$  order of resummation method in
Ref.\cite{zin-borel} and also compared to exact results. Again the Hypergeometric-Meijer algorithm in this work  gives very accurate results using a relatively low order of the perturbation sires as an input.

\begin{table}[pth]
\caption{{\protect\scriptsize {The fourth order Hypergeometric-Meijer resummation $_6F_{4}$ for
the ground state energy corresponding to the Hamiltonian in Eq.(\ref{iphi3})
compared to the $150^{th}$ order of resummation methods in
Ref.\cite{zin-borel} and also to exact results.}}}%
\label{tab:wgresum}%
\begin{tabular}
[c]{|l|l|l|l|}\hline
g & $_{\text{ }6}F_{4}$ & $150^{th}$ Order in Ref.\cite{zin-borel} & Exact\\\hline
0.5 & 0.5168918532764233 & 0.516891764253171978 & -\\\hline
1 & 0.5307847352189364 & 0.530781759304176 671 & 0.5308175930417667\\\hline
288/49 & 0.6130307602030971 & 0.612738106388984124 &
0.612738106388984125\\\hline
\end{tabular}
\end{table}
\section{Critical Exponents of the $O(4)$-symmetric quantum field model}\label{exponent}
The Lagrangian density of the $O(N)$-vector quantum field model is given by:%
\[
\mathcal{L=}\frac{1}{2}\left(  \partial\Phi\right)  ^{2}+\frac{m^{2}}{2}%
\Phi^{2}+\frac{\lambda}{4!}\Phi^{4},
\]
with  $\Phi=\left(  \phi_{1},\phi_{2},\phi_{3},...........\phi_{N}\right)  $
is an N-component field having $O(N)$ symmetry where  $\Phi^{4}=\left(
\phi_{1}^{2}+\phi_{2}^{2}+\phi_{3}^{2}+...........\phi_{N}^{2}\right)  ^{2}$. For $N=4$, it can describe the the phase transition in $QCD$  with two light
flavors at finite temperature \cite{QCD}. Recently, the six-loops order for the renormalization group functions $\beta,\gamma_{\phi^2}$ and $\gamma_{m^2}$ has been obtained in Ref.\cite{ON17}.
In Minimal-subtraction technique and in three dimensions, the six-loops order for the $\beta$-function in three dimensions is given by:
\begin{equation}
\beta(g)\approx-g+4g^2-8.667g^3+55.66g^4-533.0g^5+6318g^6-86768g^7.
\end{equation}
The large-order asymptotic behavior of this series is characterized by the parameters $\sigma=-1$ and $b=5$ \cite{Kleinert-Borel}. The strong-coupling asymptotic behavior is not yet known (up to the best of our knowledge).  The suitable weak-coupling and large order parametrized Hypergeometric-Meijer approximant for $\beta$ is then 
\begin{equation}
\beta(g)\approx-g _{\text{ }4}F_{2}(a_{1},...a_{4};b_{1}....b_{2};-g)=-g \frac{\prod_{k=1}^{2}%
\Gamma\left(  b_{k}\right)  }{\prod_{k=1}^{4}\Gamma\left(  a_{k}\right)  }%
\MeijerG*{1}{4}{4}{3}{1-a_{1}, \dots,1-a_{4}}{0,1-b_{1},  1-b_{2}}{-g},
\end{equation}
where $a_1=15.4564, a_2= -2.02503, a_3=-0.598824,a_4= -0.136248$ and $b_1=0.315181, b_2= -2.02558$. zeros of $\beta(g)$ defines fixed points where our resummation result gives $\beta(g_c)=0$ at $g=g_c=0.358732833040498$. Here $g_c$ is the critical value of the coupling where it has been predicted (but at five-loops) in Ref.\cite{keinert-strong} to be  $g_c=0.34375$. 
The critical exponent $\omega$ is defined as $\beta^\prime(g_c)$ which gives $\omega=0.7816168139530013$ compared to Borel with conformal mapping result as  0.794(9) from Ref.\cite{ON17} and Monte Carlo simulations result that gives the value $0.765$  \cite{MC11} while  the recent conformal bootstrap calculations gives  the result $\omega=0.817(30)$ \cite{Bstrab3,ON17}.

The six-loops series for the anomalous mass dimension $\gamma_{m^2}$ has been obtained in the same reference (Ref.\cite{ON17})
where:%
\begin{equation}
\gamma_{m^2}(g)\approx-2g+ 1.6667g^{2}-9.500g^{3}+64.39g^{4}-571.9g^{5}%
+5983g^{6},
\end{equation}
and the corresponding large-order parameters are $\sigma=-1$ and $b=5$. The Hypergeometric-Meijer  resummation gives the exponent $\nu$ as:
\begin{equation}
\nu^{-1}=2+\gamma_{m^2}(g_c)=2+\, _4F_2\left((a_{1},...a_{4};b_{1}....b_{2};-g_c\right)=2+\frac{\prod_{k=1}^{2}%
\Gamma\left(  b_{k}\right)  }{\prod_{k=1}^{4}\Gamma\left(  a_{k}\right)  }%
\MeijerG*{1}{4}{4}{3}{1-a_{1}, \dots,1-a_{4}}{0,1-b_{1},  1-b_{2}}{-g_c},
\end{equation}
 which gives $\nu=0.7441813061765146$. The  Monte Carlo simulations result from Ref.\cite{MC11} gives  0.750(2) and the recent Borel with conformal mapping result is 0.7397(35) \cite{ON17} while conformal bootstrap gives the result $\nu=0.751(3)$ in Ref.\cite{Bstrab3}. It is very clear that our algorithm gives very precise results.

The six-loops order for  the field anomalous dimension $\gamma_{\phi^2}$ is \cite{ON17}

\begin{equation}
\gamma_{\phi^2}(g)\approx  g^{2}(0.16667-0.16667g +0.9028g^{2}-6.5636g^{3}%
+55.93g^{4}),
\end{equation}
with $\sigma=-1$ and $b=4$ \cite{Kleinert-Borel}. The resummation result is 

\begin{equation}
\gamma_{\phi^2}(g)=0.16667 g^2_{\text{ }3}F_{1}(a_{1},...a_{1};b_{1};-g)=0.16667 g^2 \frac{\Gamma\left(  b_{1}\right)  }{\prod_{k=1}^{3}\Gamma\left(  a_{k}\right)  }%
\MeijerG*{1}{3}{3}{2}{1-a_{1}, \dots,1-a_{3}}{0,1-b_{1}}{-g}.
\end{equation}
Our resummation result gives $\eta=2 \gamma_{\phi}(g_c)=0.036694826653350686$ compared to Monte Carlo result 0.0360(3) \cite{MC11} and recent Borel with conformal mapping result 0.0366(4)\cite{ON17} while recent conformal bootstrap calculations for $\eta$ is $0.0378(32)$ \cite{Bstrab5}. The critical exponents predictions of this work is summarized in table-\ref{6LL} and compared to recent resummation results as well as simulations results.   We will not go far for such type of calculations as a full discussion of the critical exponents of the $O(N)$ model will appear in another work. 
\begin{table}[ht]
\caption{{\protect\scriptsize { The six-loops   Hypergeometric-Meijer
resummation  for the critical exponents $\nu,\eta$ and $\omega$   for $O(4)$-symmetric model. The results are compared to recent Conformal bootstrap calculations (second) Borel with conformal mapping    resummation (third)  from Ref.\cite{ON17} and also recent Monte Carlo   simulations methods (last) from Ref.\cite{MC11}.}}}%
\label{6LL}
\begin{tabular}{|l|l|l|l|l|}
\hline
N & $\ \ \ \ \ \nu$ & $\ \ \ \ \ \eta$ & $\ \ \ \ \ \omega$ & Reference \\ \hline
4 & \begin{tabular}[c]{@{}l@{}}0.74418\\ 0.751(3)\\ 0.7397(35)\\ 0.750(2)\end{tabular} & \begin{tabular}[c]{@{}l@{}}0.0366948\\0.0378(32)\\ 0.0366(4)\\ 0.0360(3)\end{tabular} & \begin{tabular}[c]{@{}l@{}}0.78162\\ 0.817(30)\\0.794(9)\\ 0.765\end{tabular} & \begin{tabular}[c]{@{}l@{}}This work\\ \ \ \ \cite{ON17,Bstrab3,Bstrab5} \\ \ \ \  \cite{ON17} \\ \ \ \  \cite{MC11}\end{tabular} \\ \hline
\end{tabular}%
\end{table}
\section{Resummation of the seven-loops $\beta$-function of the four dimensional $\phi^4$ scalar field theory}\label{beta}

In $\overline{MS}$-Scheme, the seven-loops of the perturbation series of the $\beta$-function for  the $\phi^{4}_{3+1}$ scalar field theory has been recently obtained in Ref.\cite{7L} as:
\begin{equation}
\beta\approx 3.000g^{2}-5.667g^{3}+32.55g^{4}-271.6g^{5}%
+2849g^{6}-34776 g^{7}+474651g^{8}.
\end{equation}
The $\phi^{4}_{3+1}$ theory is well known to have no fixed points and the series above has been  recently resummed using the Borel-Hypergeometric resummation algorithm \cite{HMGNB}. The results of the Borel-Hypergeometric  resummation  assured the non-existence of fixed points for the theory but on the other hand the convergence of the calculations was not perfect. We resummed the same series using our algorithm where we get:
 
\begin{equation}
\beta= 3g^2 _{\text{ }4}F_{2}(a_{1},...a_{4};b_{1}....b_{2};-g)=3g^2 \frac{\prod_{k=1}^{2}%
\Gamma\left(  b_{k}\right)  }{\prod_{k=1}^{4}\Gamma\left(  a_{k}\right)  }%
\MeijerG*{1}{4}{4}{3}{1-a_{1}, \dots,1-a_{4}}{0,1-b_{1},  1-b_{2}}{-g},
\end{equation}
To monitor the convergence of calculations and thus compare with those presented in Fig.1 of Ref.\cite{HMGNB}, we generated   the five and six loops resummations and plot all the results in Fig.\ref{Beta567}. In the figure, the calculation proves also non-existence of any fixed points for the theory but in our calculations the convergence has been greatly improved when compared to Fig.1 in  Ref.\cite {HMGNB}

\begin{figure}[H]
\begin{center}
\epsfig{file=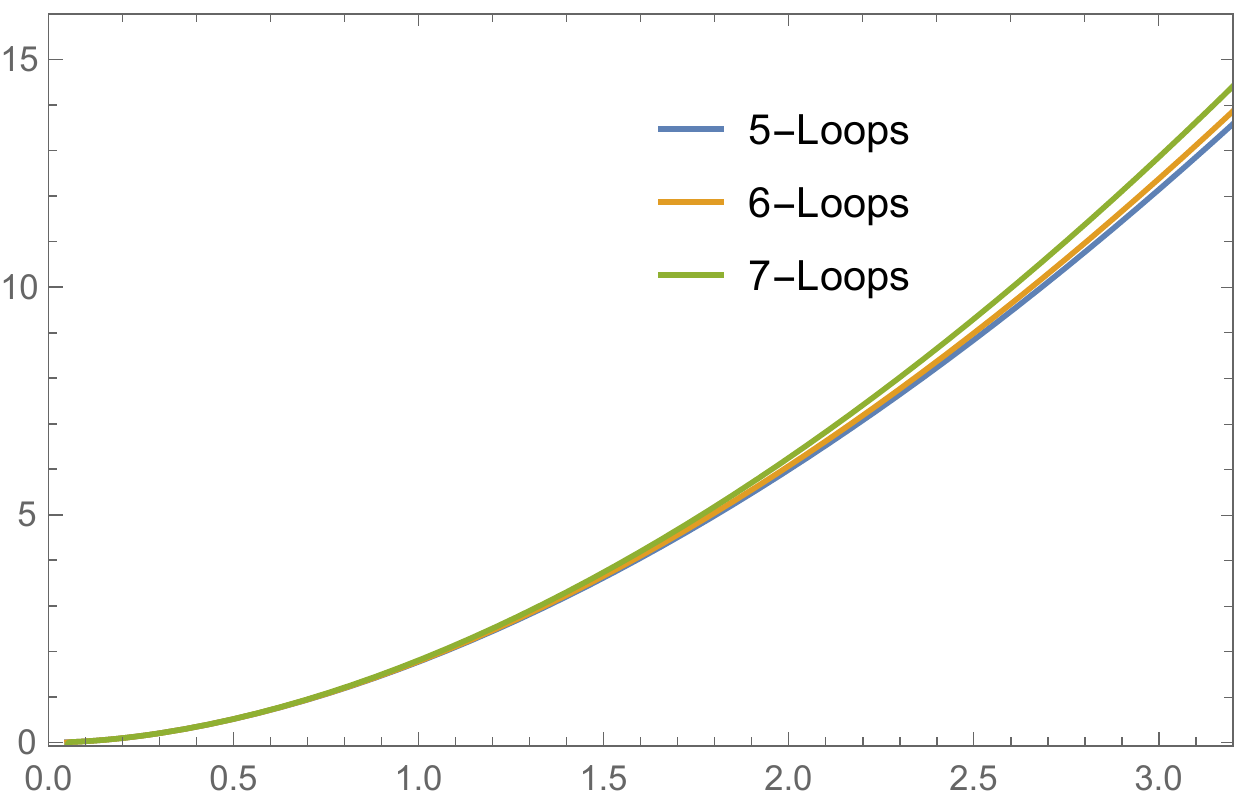,width=0.65\textwidth}
\end{center}
\caption{\textit{The Hypergeometric-Meijer resummation of the five, six and seven loops of the $\beta$-function of the four-dimensional $\phi^4$ theory.  }}%
\label{Beta567}%
\end{figure}

\section{Resummation of the seven-loop $\varepsilon$-expansion  of fractal dimension of the critical curves for the self avoiding polymer}\label{fractal}
Recently, in Ref.\cite{Fractal}, the authors obtained the six-loop $\varepsilon$ expansion  for the fractal dimension $d_f$ for the case   $N=0$ of the $O(N)$-symmetric $\phi^4$ model. They introduced what they called self-consistent resummation procedure and used it to resum the associated divergent series. However, the seven-loop $g$-expansion has been recently obtained \cite{7L} from which one can extract the seven-loop order of the $\varepsilon$-expansion for any $N$. We shall stress her  only the $N=0$ case which is in the same class of universality with the self-avoiding polymer. In fact, the resummation for the seven-loop critical exponents for different $N$ values will appear in another work\cite{AboE7}.
Using the seven-loop expansions in Ref.\cite{7L}, we obtain the flowing perturbation series up to $\varepsilon^7$ for the fractal dimension $d_f$: 
\begin{equation}
d_f(\varepsilon)=2.0000-0.25000\varepsilon-0.085938\varepsilon^{2}+0.11443\varepsilon
^{3}-0.28751\varepsilon^{4}+0.95613\varepsilon^{5}-3.8558\varepsilon
^{6}+17.784\varepsilon^{7}. \label{nueps0}%
\end{equation}
Note that the first six terms in this result are compatible with the six-loop result in Ref.\cite{Fractal}. The large order parameters are $\sigma=\frac{3}{8}$ and $b=4$ \cite{Kleinert-Borel}. We used this series to parametrize the hypergeometric approximant $2\  _{5}F_{3}%
(a_{1},...a_{5};b_{1},b_{2},b_{3};-\sigma\varepsilon)$ which in turn leads to the  result:
\begin{equation}
d_f\approx\frac{2\Gamma(b_{1})\Gamma(b_{2})\Gamma(b_{3})}{\Gamma\left(
a_{1}\right)  \Gamma\left(  a_{2}\right)  \Gamma\left(  a_{3}\right)
\Gamma\left(  a_{4}\right)  \Gamma\left(  a_{5}\right)  }\MeijerG*{1}{5}{5}{4}{1-a_{1},1-a_{2},1-a_{3},1-a_{4},1-a_{5}}{0,1-b_{1}%
,1-b_{2},1-b_{3}}{-\frac{3}{8}\varepsilon}
\end{equation}
This approximant yields the result $d_f=1.3307$  for the two dimensional ($\varepsilon=2$) self-avoiding polymer. Note that the conformal field theory (also exact) result is $\frac{4}{3}=1.33333$ \cite{Fractal,CFT1,CFT2,CFT3,exact} while the recent self-consistent resummation result is $1.354(5)$. It very clear that our seven-loop resummation result is very close to the exact result.

\section{Summary and Conclusions} \label{conc}

We introduced  what we can call  it the   Hypergeometric-Meijer  algorithm for a resummation of a divergent series with zero radius of convergence. The suggested algorithm  is capable of accommodating the large-order and strong coupling information and thus is able to accelerate the convergence to the exact results. In  Ref.\cite{Prd-GF}, H\'{e}ctor Mera \textit{et.al} followed a Borel-Pad$\acute{e}$ technique that led to  a Meijer-G approximant algorithm which has been shown to produce precise results from  weak coupling information as input. The algorithm we introduced however avoids Borel or Pad$\acute{e}$ techniques used in Ref.\cite{Prd-GF} and instead starting  from the parametrization of a Hypergeometric function that has the same $n!$ growth factor characterizing the divergent series and then use the equivalent integral representation of Meijer G function as an approximant to the given perturbation series. In fact, using weak coupling information in both the Hypergeometric-Meijer  G approximant in our work and that in Ref.\cite{Prd-GF} leads to different parametrization. This can be seen from the exact partition function of zero-dimensional $\phi^4$ theory which has been obtained by a third order parametrization of Hypergeometric-Meijer  G approximant in our work while in  Ref.\cite{Prd-GF} the same result has been obtained at the fifth order. 
 
Incorporation of the large-order information has been shown to accelerate the convergence and in adding the  strong  coupling data into the  resummation technique, the convergence is even faster a fact that is traditionally known in resummation techniques \cite{Kleinert-Borel}. In our work, however, we obtained a new constraint on the parameters of the Hypergeometric approximant which relates them to one of the parameters in the large-order asymptotic behaviour of the given perturbation sires. The validity of this constraint has been tested in our work by obtaining the exact result of the zero-dimensional partition function of the $\phi^{4}$ theory at the first order parametrization using weak-coupling and large-order data while in adding strong coupling data, the exact result is completely  parametrized from large-order and strong-coupling data. In both of these different parametrizations that lead to the exact result,    the constraint  ($\sum_{i=1}^{p}a_{i}-\sum_{i=1}^{p-2}b_{i}-2=b$) ~ on the parameters  has been applied.

The algorithm is also applied to resum the ground state energies of the
$\phi^{4}$  as well as the $\mathcal{PT}$-symmetric $i\phi^{3}$
field theories in $0+1$ space-time dimensions (quantum mechanics). It shows
precise predictions although few number of perturbative terms are employed.

It is well known that till now a closed form of the strong-coupling asymptotic behavior for quantum field theories in dimensions greater than one has not been obtained yet. In literature one can find that some predictions for the asymptotic behavior can be extracted from resummation techniques. Since these resummation  include free parameters that have to be optimized to give the best convergence which then lead to a prediction of the strong-coupling behavior. However, different optimizations can lead to different results for the same theory. Our algorithm on the other hand has no free parameters and thus for the same theory and same input information shall give a unique prediction for the strong-coupling behavior of a divergent series. We tested our algorithm regarding this fact and obtained an accurate prediction for the asymptotic behavior of the ground state energy of the $(\phi^4)_{0+1}$ theory using weak-coupling and large order data as input. 

In the literature one can find that  the first few orders of the epsilon expansion can lead to accurate results for critical exponents but on the other hand this accuracy is ruined by adding higher orders. An answer to that puzzle can be obtained from our resummation algorithm as if any of the numerators parameters is negative integer say $-2$, the Hypergeometric approximant is now a truncated Hypergeometric polynomial of order $2$ and thus higher orders of the perturbation series are irrelevant.  

Since the type of divergent series stressed in this work shares the same properties
of the divergent series representing the renormalization group functions in
quantum field theory \cite{Kleinert-Borel}, we applied it to resum the recent six-loops orders for the $\beta, \gamma_{m^2}$ and $\gamma_{\phi^2}$ renormalization group functions of the $O(4)$-symmetric model in three dimensions. Very precise estimation of the corresponding critical coupling as well as critical exponents have been extracted from resummation results of the renormalization group functions.

The $\phi^4$ scalar field theory is well known to have no fixed points in four dimensions. The Hypergeometric-Borel resummation algorithm has been applied recently to resum the recent seven-loops perturbative order of the $\beta$-function. The result of that algorithm asserts the non-existence of fixed points but the convergence of calculations is questionable. We resummed the same series using our algorithm where our calculations shows also no fixed points but on the other hand convergence of the calculations has been greatly improved. 

The seven-loop perturbation series ($\varepsilon$-expansion) for the fractal dimension $d_f$ of the self-avoiding polymer has been listed in this work. Resumming that series using our algorithm introduced in this work  gives a very accurate result for the two dimensional case ($\varepsilon=2$). Note that, in two dimensions, the $\varepsilon$-series is well known to have a slower convergence  than the three dimensional case and thus offers a challenging test to our resummation algorithm. The accurate result we obtained ($d_f=1.3307$) reflects an extraordinary success to our rsummation method  specially when we know that the exact value is $d_f=4/3\approx1.3333$. Our resummation result might be the most accurate resummation prediction for the same series in literature.  
 
Since the Meijer-G function is represented  by a Mellin-Barnes type of integrals, there is a possibility for the existence of Stokes phenomena  \cite{Stokes}. So one can have Hypergeometric-Meijer non-summability like cases of non-Borel summability. For such cases one resorts to the resummation of resurgent transseries which then kills the complex ambiguity \cite{Stokes1}. We applied our algorithm to resum the transeeries of the partition function of degenerate vacua $\phi^4$ theory where we obtained exact result at the third order parametrization of Hypergeometric-Meijer approximant. After incorporating the large order data, the same result has been obtained using first order parametrization.

\end{document}